\documentclass[sigconf]{acmart}
\usepackage{balance}
\usepackage{booktabs} 
\usepackage{latexsym}
\usepackage{makecell}
\usepackage{epsfig}
\usepackage{algorithm, algpseudocode}
\usepackage{booktabs}
\usepackage{multirow}
\usepackage{color}
\usepackage{subfigure}
\usepackage{mathrsfs}
\usepackage{amsmath}
\usepackage{bm}

\newcommand{\secref}[1]{Section \ref{#1}}
\newcommand{\figref}[1]{Figure \ref{#1}}
\newcommand{\eqnref}[1]{Eq. (\ref{#1})}

\newcommand{\tabref}[1]{Table \ref{#1}}
\newcommand{\bi}[1]{\textbf{\textit{#1}}}

\setcopyright{acmcopyright}

\begin{document}
	
\fancyhead{}

\title{Conceptualize and Infer User Needs in E-commerce}

%
%
%
%
%

\author{Xusheng Luo$^{1}$, Yonghua Yang$^{1}$, Kenny Q. Zhu$^{2}$, Yu Gong$^{1}$, Keping Yang$^{1}$}
\affiliation{
	\institution{$^{1}$Alibaba Group, Hangzhou, China \\$^{2}$Shanghai Jiao Tong University, Shanghai, China}
}
\email{{lxs140564, huazai.yyh, gongyu.gy, shaoyao}@alibaba-inc.com; kzhu@cs.sjtu.edu.cn}

\copyrightyear{2019} 
\acmYear{2019} 
\acmConference[CIKM '19]{The 28th ACM International Conference on Information and Knowledge Management}{November 3--7, 2019}{Beijing, China}
\acmBooktitle{The 28th ACM International Conference on Information and Knowledge Management (CIKM '19), November 3--7, 2019, Beijing, China}
\acmPrice{15.00}
\acmDOI{10.1145/3357384.3357812}
\acmISBN{978-1-4503-6976-3/19/11}

\begin{abstract}
Understanding latent user needs beneath shopping behaviors is
critical to e-commercial applications.
Without a proper definition of user needs in e-commerce,
most industry solutions are not driven directly by user needs at current stage, 
which prevents them from further improving user satisfaction.
Representing implicit user needs explicitly as nodes like ``outdoor barbecue'' 
or ``keep warm for kids'' in a knowledge graph, 
provides new imagination for various e-commerce applications.
Backed by such an e-commerce knowledge graph, 
we propose a supervised learning algorithm to conceptualize user needs from 
their transaction history as ``concept'' nodes in the graph 
and infer those concepts for each user
through a deep attentive model.
Offline experiments demonstrate the effectiveness and stability of our model,
and online industry strength tests show substantial advantages of such 
user needs understanding. 
\footnote[2]{Kenny Q. Zhu was partially supported by NSFC grant 91646205 and Alibaba visiting scholar program.}
\end{abstract}

%
%



\maketitle

\section{Introduction}
\label{sec:intro}


Intuitively, knowing what users need in their mind when they come to 
the shopping platform is vital to e-commerce giants like Alibaba and Amazon.
However, user needs in e-commerce are not well defined,
making it difficult for various e-commerce applications to truly understand their users, which gradually becomes the bottleneck to further improve user satisfaction in e-commerce.
For example, 
item recommendation, one of the major applications in e-commerce,
widely adopts the 
idea of item-based \textit{collaborative filtering (CF)} \cite{linden2003amazon, sarwar2001item}.
The recommender system uses user's historical behaviors as triggers to recall a small set of most similar items as candidates, 
then recommends items with highest weights after scoring with a ranking model. 
A critical shortcoming of this framework is that it is not driven by user needs 
in the first place, which inevitably makes
it hard for the recommender system to jump out of 
historical behaviors to explore other implicit user needs.
Besides, items recommended are hard to be explained 
except for trivial reasons such as ``similar to those items you have already viewed or purchased''.
Therefore, despite its widespread use,
the performance of current recommendation systems is still under criticism. 
Users are complaining that some recommendation results are redundant 
or lack of novelty, 
since current recommender systems can only satisfy very limited user needs 
such as the needs for a particular category or brand.
Without the ability of inferring user needs comprehensively and accurately, 
it is difficult for current systems to recommend items which a user 
may never think of but potentially have interests on, or provide convincing 
recommendation reasons to help users make shopping decisions. 

\begin{figure*}[th]
	\centering
	\epsfig{file=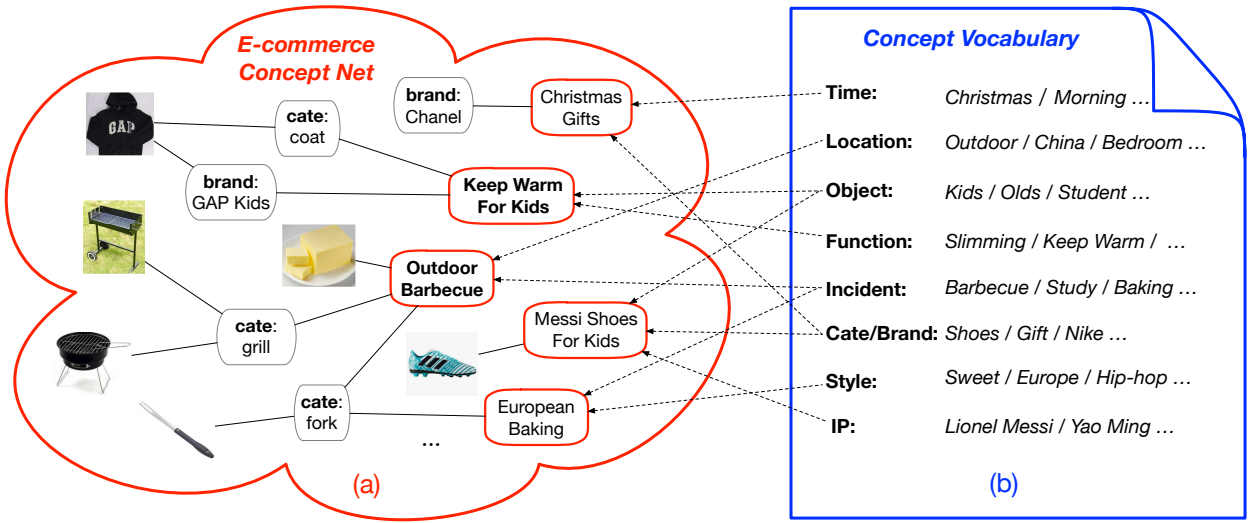, width=2\columnwidth}
	\caption{(a) Overview of ``E-commerce Concept Net'', where concepts are marked by red rectangles and pictures are  example items. (b) Overview of concept vocabulary, where each concept can be expressed using the values from eight different domains.}
	\label{fig:kg}
\end{figure*}

In this paper, we attempt to conceptualize various implicit user needs in e-commerce 
scenarios as explicit nodes in a knowledge graph, 
then infer those needs for each user.
By doing that, our platform is able to suggest a customer 
``other items you will need for outdoor barbecue next week'' 
after he purchases a grill and clicks on charcoals,
or remind him of preparing clothes, hats or scarfs that can ``keep warm for your kids'' as 
there will be a snowstorm coming next week.
Different from most e-commerce knowledge graphs, which only contain
nodes such as categories or brands,
a new type of node, e.g.,
``Outdoor Barbecue'' and ``Keep Warm for kids'', is introduced as 
bridging concepts connecting user and items to satisfy some 
high-level user needs or shopping scenarios. 
We call these nodes ``\textbf{e-commerce concepts}'',
whose structure represents a set of items from different categories 
with certain constraints (more details in \secref{sec:ecn}) .
These e-commerce concepts, together with categories, brands and items, 
form a new kind of e-commerce knowledge graph, called
\textbf{``E-commerce Concept Net''} (\figref{fig:kg} (a)). 
For example, ``Outdoor Barbecue'' is one such e-commerce concept,  
consisting of product categories such as charcoal, forks and so on, 
which are items required to host a successful outdoor barbecue party.

\begin{figure}[th]
	\centering
	\epsfig{file=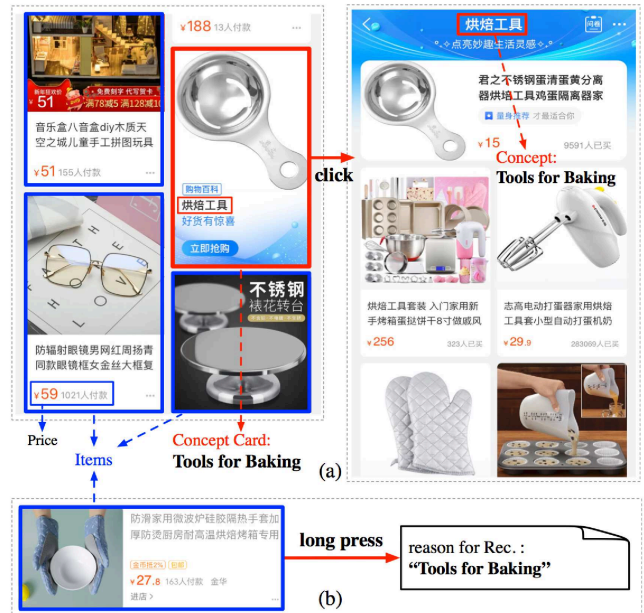, width=\columnwidth}
	\caption{Two real examples of user-needs driven recommendation.
		(a) Display concepts directly to users as cards with a set of related items. (b) Concepts act as explanations in item recommendation.
	}
	\label{fig:cloud}
\end{figure}

There are several possible practical scenarios in which 
inference of such e-commerce concepts from user behaviors can be useful. 
The first scenario is coarse-grained recommendation,
where inferred concepts can be directly recommended to users 
together with its associated items.
\figref{fig:cloud}(a) shows the real implementation of this idea in 
\textit{Taobao} \footnote{\url{http://www.taobao.com}} App.
Among normal recommended items, 
concept ``Tools for Baking'' is displayed to users as a card with its name and the picture of a representative item (left).
Once a user clicks on it, he will enter into another page (right) where different 
items needed for baking are displayed.
In this way, the recommender system is acting like a salesperson in a shopping mall, 
who tries to guess the needs of his customer and and then suggests how to satisfy
them. If their needs are correctly inferred, users are more likely to accept 
the recommended items.
The second scenario is providing explanations for item recommendation as 
shown in \figref{fig:cloud}(b).
While explainable recommendation attracts much research attention 
recently \cite{zhang2018explainable}, 
most existing works are not practical enough for industry systems,
since they are either too complicated 
(based on NLG \cite{zanker2010knowledgeable,cleger2012explaining}), or too trivial 
(e.g., ``how many people also viewed'' \cite{costa2018automatic,li2017neural}).
Our proposed concepts, on the contrary, precisely conceptualize user needs and are
easy to understand. This idea is currently experimented in Taobao at the time of 
writing. Other possible scenarios can be query rewriting or query suggestion in e-commerce search engine.

User needs inference backed by a knowledge graph (KG) is a relatively new problem. 
The most related work is incorporating KG into recommendation~\cite{zhang2016collaborative,sun2018recurrent,huang2018improving}.
Prior efforts are mainly categorized into two types. 
Path based methods \cite{zhao2017meta,hu2018leveraging} explore the various patterns of connections among items in KG, providing rich \textit{meta-path} based features for
user-item recommendations.
Those methods generally treat KG as a heterogeneous information network (HIN) and rely on manually crafted meta-paths.
The other line of research \cite{wang2018dkn,huang2018improving} leverage knowledge graph embedding (KGE) such as TransE \cite{bordes2013translating}, to bring extra information from KG to enhance the representation of items and users. 
However, KGE based methods usually lack the ability to reason across multiple 
hops and have not shown to be scalable on large-scale dataset.
Different from most existing works targeting item (or movie/news), 
the target (concept) in our problem is a set of items, which itself has a non-trivial structure and contains much more information than a single item.
In order to handle the informative input and provide more interpretability,
we further extend the direction of path based works
by proposing a deep interpretable model with a specially designed module 
called ``attention cube'', which
aims to explore the mutual influences among users, concepts and paths 
connecting user-concept pairs within the concept net.


The contributions of this paper are summarized below:
\begin{itemize}
	\itemsep0em
	\item We formally define user needs in e-commerce and introduce ``e-commerce concept net'', a new genre of knowledge graph in e-commerce, where ``concepts'' can explicitly express various shopping needs for users. 
	\item Based on the e-commerce concept net,
	we propose a path-based deep model with attention cube to infer user needs. 
	We evaluate our model in both offline and online settings.
	Offline results show the model outperforms 
	several strong baselines by a substantial margin of $\textbf{2.4\%}$ on AUC.
	Online testing deployed on a real recommender system in Taobao also achieves largest improvement on CTR and Discovery. $\textbf{20.5\%}$ improvements on User Satisfaction Rate further indicates the value of such user needs inference.
	\item Our model has already gone into production of Taobao, the largest e-commerce platform in China. We believe the idea of user needs understanding can be further applied in more e-commerce productions.
	There is ample room for imagination and further innovation in ``user-needs driven'' e-commerce.
\end{itemize}


\section{E-commerce Concept Net} 
\label{sec:ecn}

User needs in e-commerce, are not formally defined previously.
Hierarchical categories and browse nodes \footnote{\url{https://www.browsenodes.com/}} are ways of managing billions of items in e-commerce platforms
and are usually used to represent user needs or interests \cite{zhou2018deep, feng2019deep}.
However, user needs are far broader than categories or browse nodes. Imaging a user who is planning an outdoor barbecue, or who is concerned with how to get rid of a raccoon in his garden.
They have a situation or problem but do not know what products can help.
Therefore, tree-like structures such as hierarchical categories and browse nodes are not enough to represent those user needs.

In our e-commerce concept net \footnote{This section only gives
a brief introduction of the e-commerce concept net, while more details will be 
discussed in a separate paper which will be released in the near future at \url{https://github.com/angrymidiao/concept_net}.},
user needs are conceptualized as various shopping scenarios, also known as ``e-commerce concepts''.
We \textbf{define} a proper concept being a short, fluent and reasonable phrase which naturally represents a set of items from different categories.
In order to cover as many user needs as possible,
a thorough analysis on query logs, product titles and other e-commercial text is conducted.
Based on years of experience in e-commerce,
each concept is expressed using values drawn from $8$ different domains of
an ``e-commerce concept vocabulary'', which is shown in \figref{fig:kg} (b).
For example, ``Outdoor Barbecue'' can be written as 
``\textit{Location}: outdoor, \textit{Incident}: barbecue'', 
and ``Breakfast for Pregnancy'' can be written as ``\textit{Object}: pregnant women, \textit{Cate/Brand}: breakfast''.

To form the complete e-commerce concept net,
concepts are related to their representative items, 
categories, brands respectively, mainly adopting the idea of semantic matching \cite{huang2013learning, shen2014learning}.
It should be noticed that there is a hierarchy within each domain. For example, ``Shanghai'' is a city in ``China'' in the domain of \textit{Location} and ``pregnancy'' is a special stage of a ``woman'' in the domain of \textit{Object}.  Vocabulary terms at different levels can be combined and result in different concepts.
Accordingly, those concepts are naturally related to form a hierarchy as well.
Besides the vocabularies to describe concepts, there are constraints to each concept. 
The aspects of concept \textit{schema} include
 \textit{gender}, \textit{life stage} \footnote{Life stage is divided into: pregnancy, infant, kindergarten, primary school, middle school and high school in Taobao.}, etc.
which actually corresponds to user profile.
For example, the schema of ``Breakfast for Pregnancy'' will be ``\textit{gender}: female, \textit{life stage}: pregnancy'', which indicates the group of users who are most likely to need this concept.

\begin{table}[th]
	\centering
	\small
	\begin{tabular}{|l|r|r|r|r|}
		\hline
		\multirow{4}{*}{Ontology Vocab.} 
		&\# Time &\# Location &\# Object &\# Func.  \\
		\cline{2-5}
		& 127 & 7,052 & 247 & 3,693 \\
		\cline{2-5}
		&\# Inci. & \# Cate/Bra. & \# Style &\# IP  \\
		\cline{2-5}
		& 9,884 & 44,860 & 1,182 & 21,230 \\
		\hline
		\# Concepts (Raw) & \multicolumn{1}{c|}{35,211} &
		\multicolumn{2}{c|}{\# Concepts (Online)} & \multicolumn{1}{c|}{7,461} \\ 
		\hline
		\# Items & \multicolumn{1}{c|}{1 billion} &
		\multicolumn{2}{c|}{\# Categories/Brands} & \multicolumn{1}{c|}{19K/5.5M} \\ 
		\hline
	\end{tabular}
	\caption{Statistics of E-commerce Concept Net.}
	\label{tab:data}
\end{table}

\tabref{tab:data} shows the statistics of the concept net used in this
paper~\footnote{Preview of concept data can be found at \url{https://github.com/angrymidiao/concept_net}.}.
There are 35,211 concepts in total at current stage, 
among which 7,461 concepts are already deployed in our online recommender system, covering over 90\% categories of Taobao and each concept is related with 10.4 categories on average.

Inspired by the construction of open-domain KGs such as Freebase \cite{bollacker2008freebase} and DBpedia \cite{auer2007dbpedia} which benefit various downstream applications \cite{luo2018knowledge,luo2018cross}, different kinds of KGs in e-commerce are constructed
to describe relations among users, items and item attributes \cite{catherine2017explainable,ai2018learning,gong2019deep}.
One famous example is the ``Product Knowledge Graph'' \footnote{\url{https://blog.aboutamazon.com/innovation/making-search-easier}} of Amazon. Their KG mainly supports semantic search, aiming to help users search for products that fit their need with search queries like ``items for picnic''. The major difference is that they never conceptualize user needs as explicit nodes in KG as we do.
In comparison, our e-commerce concept net introduces a new node to explicitly represent user needs. Besides, it becomes possible to link our e-commerce KG to open-domain KGs through the concept vocabulary, 
making our concept net even more powerful.

\section{Problem}
\label{sec:problem}

In this section, we formally define the problem of user needs inference.
Let $\bi{U}$, $\bi{V}$ denote the sets of users, items respectively.
The inputs of our problem are as follows:

\noindent
\textbf{1) User behavior on items}. For each $u\in \bi{U}$,  a behavior sequence 
$b= \{b_1, b_2, \cdots, b_n\}$ is a list of behaviors in time order, 
where $b_i$ is the $i^{th}$ behavior and $b_n$ is the latest one. 
Each user behavior contains a user-item interaction, 
detailed as $b_i = <v_i, type_i, time_i>$, where $v_i \in \bi{V}$, 
$type_i$ is the type of behavior, such as click or purchase, and
$time_i$ denotes the specific time of the behavior.

\noindent
\textbf{2) E-commerce concept net}. Concept net $\bi{G}$ consists of massive triples $(h, r, t)$, 
where $h, t\in \bi{E}$, $r\in \bi{R}$ denote the head, tail and relation.
$\bi{E}$ and $\bi{R}$ are entities and relations in the concept net.
While most items in $\bi{V}$ can be linked to entities in $\bi{E}$, 
some items may not, since the item pool in e-commerce platforms changes frequently. 
The set of all concepts in $\bi{G}$ is denoted as $\bi{C}$.

\noindent
\textbf{3) Side information}. 
For each user $u\in \bi{U}$, we have corresponding profile information $h$, 
such as \textit{gender}, \textit{kid's life stage} and long-term preferred categories, etc.
For each concept $c\in \bi{C}$, we have its schema $s$ introduced in \secref{sec:ecn};

Given above inputs, the goal of user needs inference is to predict potential need in concept $c$ for each user $u$. We aim to learn a prediction function $\hat y_{uc} = \bi{F}(u, c; \theta)$, denoting the probability concept $c$ is needed by user $u$, and $\theta$ is the model parameters.

\section{Approach}
\label{sec:model}
\begin{figure*}[th]
	\centering
	\epsfig{file=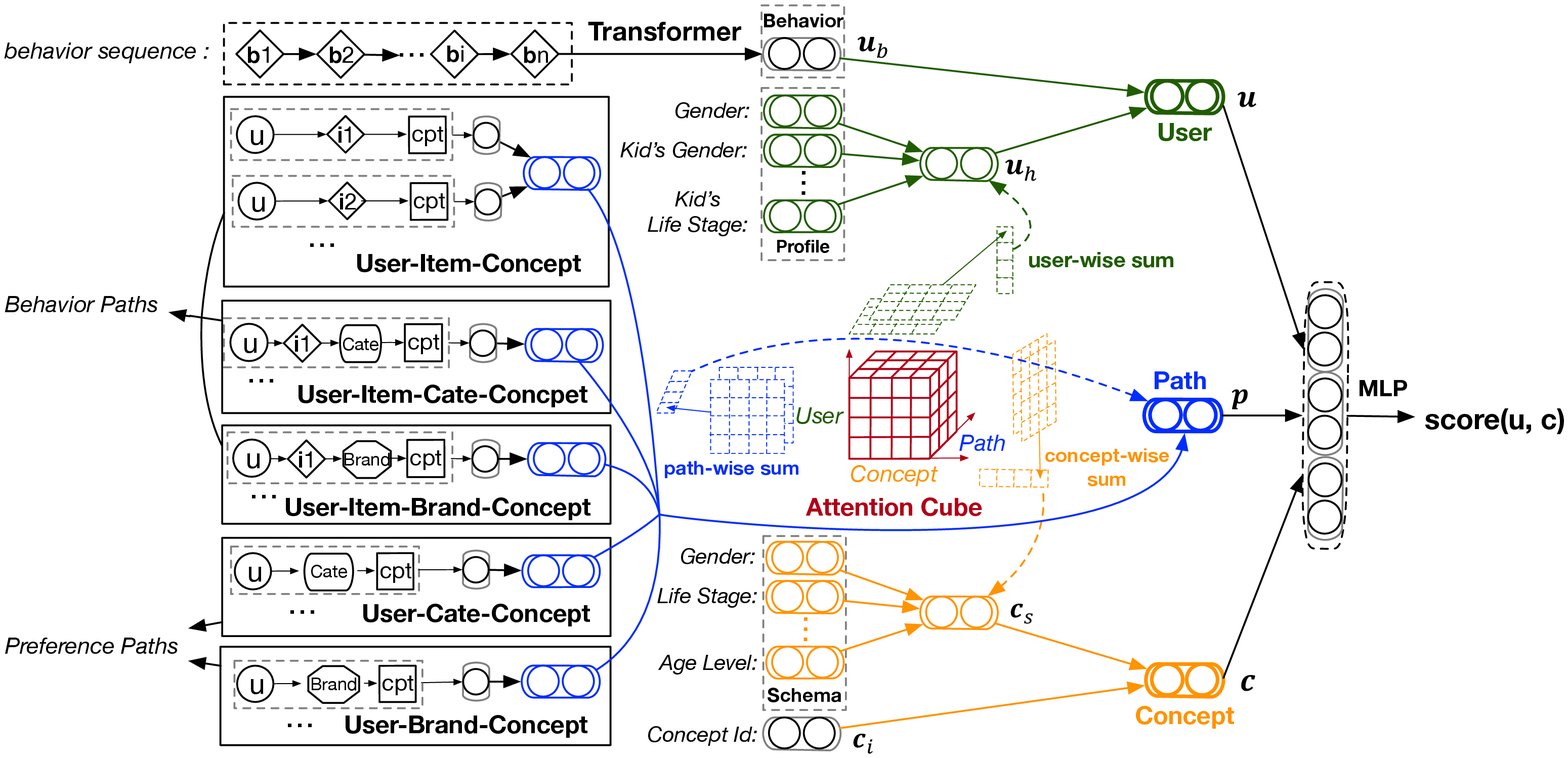, width=2\columnwidth}
	\caption{Overview of proposed model.}
	\label{fig:model}
\end{figure*}

\figref{fig:model} gives an overview of the proposed model,
which is a three-way architecture: a user, a candidate concept, and paths from the user to the concept.
Given a user and a candidate concept, 
the model leverages rich features extracted from user behavior and profile, 
candidate concept schema and path context, then outputs a score, representing the probability of the user needs the candidate concept. 

\subsection{User Embedding}

The representation for each user comes from two parts: user behavior sequence and user profile.

\noindent
\textbf{User Behavior Sequence}

\begin{figure}[th]
	\centering
	\epsfig{file=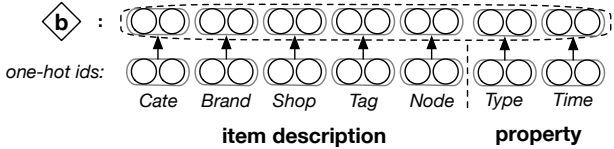, width=\columnwidth}
	\caption{Encoding of user behavior}
	\label{fig:detail_2}
\end{figure}

\noindent
Each behavior consists of three things: the item, the behavior type and the behavior time.
Due to enormous amount of items (over 1 billion) in e-commerce platform, 
we represent each item in behavior sequence 
using its description such as category, brand and shop, instead of directly using its id.
This is for two reasons: to save memory for storing large amount of id embeddings and to avoid sparsity problem when encountering long-tail or new items while predicting.
We consider four types of behavior: \textit{click}, \textit{bookmark}, \textit{add to cart} and \textit{purchase}.
In addition, the day gap between the behavior and current time is also taken into account. 
Therefore, each behavior $b_i$ can be represented as a multi-hot vector 
$[\bi{b}_{i}^{1}, \bi{b}_{i}^{2}, \cdots, \bi{b}_{i}^{F}]$,
where each one-hot vector $\bi{b}_{i}^{f}$ corresponds to one of the above mentioned feature and $F$ is the total number.
Then an embedding lookup layer shown in \figref{fig:detail_2} maps
sparse behavior vector into a low-dimensional dense vector  $\bi{b}_i$:

\begin{equation}
\label{eqn:lookup}
\bi{b}_i = [\bi{W}_{lk}^{1}\bi{b}_{i}^{1}; \bi{W}_{lk}^{2}\bi{b}_{i}^{2}; \cdots; \bi{W}_{lk}^{F}\bi{b}_{i}^{F}],
\bi{W}_{lk}^{f} \in \bi{R}^{d^{f} \times V^{f}} 
\end{equation}

\noindent
where $\bi{W}_{lk}^{f}$ are parameters for embedding lookup layer,
$d^{f}$ is the dimension of dense vector and $V^{f}$ is the vocabulary size.

Recurrent neural Networks (RNN) based models \cite{hochreiter1997long,cho2014properties} assume a rigidly ordered sequence over data which
is not always true for user behaviors in real-world applications such as e-commerce.
Such left-to-right architectures may restrict the power of the historical sequence representations. 
Thus, we believe bidirectional
model such as Transformer \cite{vaswani2017attention} with self-attention architecture is a more reasonable choice for modeling user behavior sequences.	
The embedding of user behavior sequence $\bi{u}_b$ is calculated as:
\begin{equation}
\bi{u}_b = \textbf{Transformer}(\bi{b}_1, \bi{b}_2, \cdots, \bi{b}_n)
\end{equation}

\noindent
\textbf{User Profile}

\noindent
The aspects of user profile include \textit{gender}, \textit{age level}, \textit{kid's gender}, \textit{kid's life stage}, etc.
We use a simple lookup layer similar to \eqnref{eqn:lookup} to obtain the corresponding embedding for each profile aspect.
Then we apply a function $f_{u}$ to map the embedding list 
$u_h = [\bi{h}_{\text{gender}}, \bi{h}_{\text{age}}, \cdots]$ to a single vector as the representation of user profile $\bi{u}_p$:
\begin{equation}
\label{eqn:fu}
\bi{u}_h = f_u(u_h) = f_u([\bi{h}_{\text{gender}}, \bi{h}_{\text{age}}, \cdots])
\end{equation}
where the simplest $f_u$ is average pooling. Optimizations for $f_u$ will be discussed in \secref{sec:att}.

Finally we get the user embedding $\bi{u}$ by concatenation plus a fully connected layer:
\begin{equation}
\bi{u} = \textbf{FC}([\bi{u}_{b};\bi{u}_{h}])
\end{equation}

\subsection{Concept Embedding}

Similar to user embedding which comes from user behavior and user profile, we use two components to encode the candidate concept: concept id and concept schema.
The representation of concept id $\bi{c}_i$ is obtained simply by lookup.
For concept schema,
we use embedding lookup layer to map one-hot vectors (aspects of concept schema) to dense vectors, then apply function $f_{c}$ to obtain
the representation of the concept schema $\bi{c}_s$. 
Similar to the encoding of user profile,
we then obtain concept embedding $\bi{c}$ :
\begin{equation}
\label{eqn:fc}
\bi{c}_s = f_c(c_s) = f_c([\bi{s}_{\text{gender}}, \bi{s}_{\text{age}}, \cdots])
\end{equation}
\begin{equation}
\bi{c} = \textbf{FC}([\bi{c}_{i};\bi{c}_{s}])
\end{equation}

\subsection{Path Embedding}
\label{sec:path}
In order to leverage rich semantic features from the e-commerce concept net,
we explore paths connecting users and concepts within the graph.
We adopt the idea of \textit{meta-path} \cite{hu2018leveraging}, due to the fact KGs in e-commerce are usually extremely large.
If we let the model discover possible paths from a behaved item to a concept freely as described in RippleNet \cite{wang2018ripplenet}, 
the computational overhead is unacceptable.
Besides, empirical experience is valuable in e-commerce.
Therefore, we believe manually crafted meta-paths are able to reduce noises and improve efficiency.
A meta-path is a path in the form of ``$T_1 \rightarrow T_2 \rightarrow \cdots T_n$'', where each node (exclude user in our case) is a type of entity in the concept net, such as ``$User \rightarrow Item \rightarrow Category \rightarrow Concept$''.
We mainly consider two types of meta-path in our concept net: \textit{behavior} path and \textit{preference} path.
Behavior paths are triggered by items which a user clicks or purchases, such as ``\textbf{UIC}'' (User-Item-Concept), and  ``\textbf{UITC}''(``\textbf{T}'' for ``Ca\textbf{T}egory'').
Preference paths are triggered by long-term preferred categories or brands, such as ``\textbf{UBC}''(``\textbf{B}'' for ``\textbf{B}rand'').

Within each meta-path, 
there are multiple specific paths called \textit{path instance}s.
For each meta-path, we sample a fixed number of path instances with highest priority scores.
Calculation of the priority score for each edge in a path instance is based on heuristics.
In the concept net, one item may belong to several concepts, while each concept also contains many items. So we mainly adopt \textit{tf-idf} score to measure the importance of each ``item-concept'' edge and other types of edge.
The score of the whole path instance is then calculated as the product of all the edge scores.
Then we use a Convolution Neural Network (CNN) to encode each sampled instance $\bi{pi}$ and followed by a max-pooling operation to get the embedding of that meta-path
(take ``\textbf{UITC}'' as an example):
\begin{equation}
\label{eqn:cnn}
\bi{pi}_{\text{UITC}} =  \textbf{CNN}([\bi{u}_b, \bi{i}, \bi{cate}, \bi{c}_i])
\end{equation}
\begin{equation}
\bi{p}_{\text{UITC}} = \textbf{MaxPooling}(\{\bi{pi}_{\text{UITC}}\})
\end{equation}
where $\bi{i}$ is the item embedding which only uses item description, 
and $\bi{cate}$ is id lookup embedding of category. As for head and tail node, $\bi{u}_b$ is the behavior embedding and $\bi{c}_i$ is the lookup embedding of concept id.
Comparing to RNN, CNN is much faster dealing with large amount of data and able to extract sequence dependency when sequence length is relatively short.
Then the representation of meta-path context is calculated as:
\begin{equation}
\label{eqn:fp}
\bi{p} = f_p(p) = f_p(\bi{p}_{\text{UIC}}, \bi{p}_{\text{UITC}}, \cdots)
\end{equation}

\subsection{The Whole Model}

After getting the embedding for the user, the candidate concept and the paths connecting them, we concatenate the three embeddings and feed it into a MLP and the final output indicates the probability user $u$ will need concept $c$:
\begin{equation}
\hat y_{uc} = \textbf{MLP}([\bi{u};\bi{p};\bi{c}]),
\end{equation}
where the MLP module consists of two hidden layers with ReLU activation function and an output layer with sigmoid function.

We interpret user needs inference as a binary classification problem, 
where an observed user-concept interaction is assigned with a target value $1$, otherwise $0$. 
We use point-wise learning with the negative log-likelihood objective function to learn the parameters of our model:
\begin{equation}
\mathscr{L} = -\sum_{(u, c)\in D^+}{\log \hat y_{uc}} + \sum_{(u, c)\in D^-}{\log (1-\hat y_{uc})}
\end{equation}
where $D^+$ and $D^-$ are the positive and negative user-concept interaction pairs.

\subsection{Attention Mechanism}
\label{sec:att}

If we define $f_u$, $f_c$ and $f_p$ as average pooling functions, each element contributes equally all the time.
It is obviously suboptimal since different meta-paths are likely to effect users' decision making differently.
Even for the same user, the preference on the same path may change targeting different concepts.
Similarly, different aspects of user profile and concept schema can contribute to the final decision differently as well.

Attention mechanism has been been widely used to handle weighted sum of embeddings 
in recent years \cite{bahdanau2014neural,yin2016abcnn}.
We proposed a novel attention module called ``\textbf{Attention Cube}'' to 
model the mutual influence of a three way interaction simultaneously in our problem.
Attention cube is a three-dimensional tensor with $x$, $y$, $z$ axis corresponding to $user$, $path$ and $concept$. 
We extend Luong's attention equation \cite{luong2015effective} to three-dimension and define the values of attention cube $\bi{Att}$ as below:
\begin{equation}
\label{eqn:att}
att_{i,j,k} = {{u_h}_i}^T \bm{W_1} p_j + {p_j}^T \bm{W_2} {c_s}_k + {{u_h}_i}^T \bm{W_3} {c_s}_k  
\end{equation}
where ${u_h}_i$ is $i^{th}$ embedding of user profile embedding list,
 ${c_s}_k$ is $k^{th}$ embedding of concept schema embedding list,
 and $p_j$ is $j^{th}$ embedding of meta-path embedding list.
 $\bm{W_1}$,  $\bm{W_2}$,  $\bm{W_3}$ are parameter matrices.
 
Then the weights of user profile aspects, concept schema aspects and different meta-paths are obtained by first calculating axis-wise sum and then normalization:
\begin{equation}
{\bm{\alpha_u}}_i = \frac{\exp(\sum_{j}\sum_{k} att_{i,j,k})}{\sum_{i}exp(\sum_{j}\sum_{k} att_{i,j,k})}
\end{equation}

We can get ${\bm{\alpha_p}}_j$ and ${\bm{\alpha_c}}_k$ in a similar way.
Finally, the mapping functions $f_u$ (similar for $f_c$ and $f_p$) are defined to get $\bi{u}_h$ (similar for $\bi{c}_s$ and $\bi{p}$) as below:
\begin{equation}
\bi{u}_h = f_u(u_h) = {\bm{\alpha_u}}_1 \bi{h}_{\text{gender}} + {\bm{\alpha_u}}_2 \bi{h}_{\text{age}} + \cdots 
\end{equation}

Since the attention weights ${\bm{\alpha_u}}_i$, ${\bm{\alpha_p}}_j$ and ${\bm{\alpha_c}}_k$ are generated for each user-concept interaction separately, they are able to capture the complex mutual influence among the three components and result in better representations.

\section{Offline Evaluation}
\label{sec:eval}

In this section, we first introduce the dataset and experiment setup, including evaluation metrics and baselines.
Then we present the offline results and give some discussions.
Finally, we perform ablation tests to complete our experiments.

\subsection{Datasets}

Inferring e-commerce concepts a user potentially needs is a relatively new problem, 
there is no such public datasets for experiments.
To create large amounts of gold standard data to train our model, 
we collect daily log of our online system, 
where concepts are already integrated in the recommender system. 
In a module called ``Guess What You Like'' at the front page of Taobao app, concepts are displayed as cards to users among the recommended items.
There will be one concept card every ten items on average.
In the snapshot shown in \figref{fig:cloud}(a), concept ``Tools for Baking'' is displayed as a card, with the picture of a representative item.
Once users click on this card, 
it jumps to a page full of related items such as egg scrambler and strainer.
In order to alleviate the potential influence of the item picture on users' decision making,
we collect positive samples from those user-concept clicks
only if that user continues to click at least two related items after entering the concept card.
For the same reason, negative samples come from at least two exposes of the same concept (but different item pictures) without any clicks.
We collect samples for continuous four days during January 11 to January 14, 2019, and use the data of first three days for training and validation. 
We randomly select $10\%$ samples of the last day for testing.
The ratio of negative and positive is around $37:1$.
For user-item interaction data, we collect 30-days transaction records on Taobao platform for each user in our data.
Detailed statistics of our dataset is illustrated in \tabref{tab:exp_data}.
\begin{table}[th]
	\centering
	\begin{tabular}{l|c|c|c}
		\hline
		  & Training & Validation & Testing \\
		\hline
		\# of samples & 32,496,827 & 328,251 & 1,237,506 \\
		\# of users & 16,120,600 & 323,544 & 1,121,475 \\
		\# of concepts & 4,760 & 2,935 & 3,176 \\
		\hline
		\# of items & 438M & 76M  & 141M \\
		\# of categories & 15,257 & 11,799 & 14,590 \\
		\# of brands & 1,434,659 & 428,036 & 1,088,480 \\
		\hline
	\end{tabular}
	\caption{Statistics of Taobao's dataset.}
	\label{tab:exp_data}
\end{table}

Based on years of e-commerce experience, we mainly select five meta-paths (\figref{fig:model}) in our experiments: ``\textbf{UIC}'', ``\textbf{UITC}'' and ``\textbf{UIBC}'' for behavior  paths; ``\textbf{UTC}'' and ``\textbf{UBC}'' for preference paths.
Longer paths are not selected since they are likely to bring noises.

\subsection{Experiment Setup}

\noindent
\textbf{Evaluation Metrics}

\noindent
We perform evaluation of different models in two experiment scenarios.
1) In click-through-rate (CTR) prediction, we apply the trained model to each sample of test set and calculate $AUC$ based on the output score to evaluate the overall performance; 2) In top-$N$ recommendation scenario, 
we use the trained model to select $N$ concepts with highest predicted scores for each user in the test set. 
we evaluate the results by Hit Ratio ($HR@N$), and Normalized  Discounted Cumulative Gain ($NDCG@N$),
which are widely used in recommendation tasks having very few ground-truth results \cite{huang2018improving,chen2018sequential}.
In order to make sense under the second scenario, 
we augment the test set mentioned above by removing samples where the user does not have any positive clicks,
and report averaged $HR$ and $NDCG$ across users.

\noindent
\textbf{Baselines}

\noindent
We compare with the following baselines:
\begin{itemize}
	\itemsep0em
	\item \textbf{BPR} \cite{rendle2009bpr} is the Bayesian Personalized Ranking model that minimize the pairwise ranking loss for implicit feedback.
	\item \textbf{Wide\&Deep} \cite{cheng2016wide} is the widely used recommendation framework, which jointly trains wide linear models and deep neural networks. We use embeddings of users, concepts and other entities to feed Wide\&Deep.
	\item \textbf{MCRec+} is based on MCRec\cite{hu2018leveraging}, which is a state-of-the-art HIN based model for recommendation. It treats the KG as HIN and extracts meta-path based features for modeling user-target interaction. 
	We feed e-commerce concept net as the HIN for MCRec. For fair comparison, extra information appeared in our problem such as sequential user behaviors, user profile and concept schema, 
	are also fed into MCRec in a compatible way.
	\item \textbf{KPRN+} is based on KPRN \cite{wang2018explainable}, another state-of-the-art knowledge-aware recommendation model, which aims to reason over KG by composing both entities and relations. Similar to MCRec+, we feed extra information to KPRN to get KPRN+.
\end{itemize}

\noindent
\textbf{Implementation Details}

\noindent
We implement our model using the python library of TensorFlow \footnote{\url{www.tensorflow.org}}. 
We set the length of user behavior sequence to $15$,
and sampled path instance within each meta-path to at most $50$.
The dimension of entity embeddings (item, concept, category, etc.) is set to $20$, 
and the dimension of output layer is set to $32$.
The hidden state size of GRU is set to $40$.
All parameters are randomly initialized with Gaussian distribution.
We perform a mini-batch log-likelihood loss training with a batch size of $512$ for $5$ training epochs.
We use Adam optimizer \cite{kingma2014adam}, and the learning rate is initialized to $0.001$.
For all the comparison models, 
we refer to their original papers and tune the parameters using the validation set as well.
With the help of a powerful distributed TensorFlow machine learning system in Taobao, 
we use $4$ parameter servers and $20$ workers,
and the whole training process can be finished in $4$ hours.

\subsection{Results}
\label{sec:off_eval}

\begin{figure}[th]
	\centering
	\epsfig{file=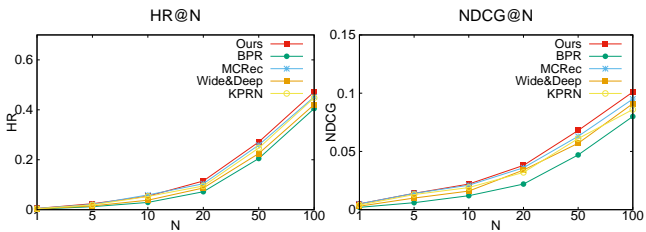, width=\columnwidth}
	\caption{HR and NDCG in Top-N recommendation.}
	\label{fig:topn}
\end{figure}

We report the experimental results in \tabref{tab:eval_main} and \figref{fig:topn}.
Our model outperforms all the baselines, improving the result by up to 2.4\% in $AUC$.
Improvements in $HR$ and $NDCG$ also reveal the superiority of our model.
BPR and Wide\&Deep perform comparably poorly than other baselines, since they do not incorporate extra knowledge from e-commerce concept net into the model, failing to leverage rich features from paths between users and concepts.
For knowledge-aware baselines, 
the main difference is the encoding of paths and attention mechanism.
MCRec+ performs best among all baselines, 
since it also try to characterize a three-way interactions among user, paths and the concept.
Our model substantially outperforms MCRec+ to achieve the best performance,
which indicates the importance of modeling mutual attentive influence of three components simultaneously.
KPRN+ performs worse than MCRec+, since the relation name matters in their problem is relatively trivial in our concept net.
The last two lines of \tabref{tab:eval_main} further demonstrate the effectiveness of our proposed attention module. 
By comparing to a degenerated version of our model, which replaces attention cube with average pooling in each component, 
our full model achieves better performance.

\begin{table}[th]
	\centering
	\begin{tabular}{l|c}
		\hline
		Model &   AUC \\
		\hline
		BPR  &  0.6005 \\
		Wide\&Deep  &  0.6137 \\
		MCRec+  &  0.6447 \\
		KPRN+  &  0.6417 \\
		\hline
		Ours (- att. cube) & 0.6403 \\ 
		Ours (full) &  \textbf{0.6612} \\
		\hline
	\end{tabular}
	\caption{AUC in CTR prediction on Taobao's dataset.}
	\label{tab:eval_main}
\end{table}

\subsection{Ablation Study}
In this subsection, we explore the contribution of various components of our model.
We report AUC on evaluation set to compare different variations in \tabref{tab:ablation}.

\noindent
\textbf{Behavior Paths vs Preference Paths}

\noindent
We first evaluate how different types of meta-path between users and concepts effect final performance. 
If we remove all paths, AUC drops by $6.8\%$, 
revealing the huge benefits brought by the concept net.
Between behavior paths and preference paths,
we can observe that AUC drops more severely when removing the former ones,
which indicates that behavior paths are more important than preference paths in our model.
It appears that recent clicks or purchases of items play a larger role 
in reflecting user needs than long-term preferences,
which may inflect that user needs are changeable and unstable, and they can be easily influenced.
\begin{table}[th]
	\centering
	\begin{tabular}{l|c|c}
		\hline
		Variation  & AUC   & Decrease (\%)\\
		\hline
		- behavior paths  &  0.6826 & 4.03\\
		- preference paths  & 0.6934 & 2.41\\
		- all paths & 0.6694 & 6.08\\
		\hline
		\hline
		- user behavior sequence & 0.7010 & 1.30\\
		- user profile & 0.6986 &  1.65\\
		- concept schema & 0.7031 & 1.00\\
		\hline
		\hline
		Full  &  \textbf{0.7101} & 0.0 \\ 
		\hline
	\end{tabular}
	\caption{Ablation tests on validation set.}
	\label{tab:ablation}
\end{table}

\noindent
\textbf{Behavior Sequence vs Side Information}

\noindent
Now we investigate the influence of user behavior sequence and side information in our problem,
where side information further includes user profile and concept schema.
Ablation towards these three components shows that they all contributes to the final inference result, 
while user profile information matters most ($1.65\%$ decrease in AUC).
It is observed that user profile seems more important than user behavior sequence.
The possible reason is that the attention cube degenerates to a matrix, if we remove user profile from the model.
This may lead to a decrease in final performance.

\section{Online Application}

The above offline experimental results have shown superiority of our proposed model for accurately inferring user needs.
Now, we deploy our model online and integrate it into a recommender system in Taobao with a standard A/B testing configuration to answer the following three questions:
\begin{enumerate}
	\itemsep0em
	\item Does our inference model still perform the best at online setting regarding both accuracy and novelty?
	\item Does user needs inference actually improve user satisfaction?
	\item Comparing to traditional item recommendation, does user-needs driven recommendation with concept cards bring extra value to e-commerce platforms? 
	
\end{enumerate}

\subsection{Experiment Setup}
The experiments are conducted in the online module introduced in \figref{fig:cloud} (a).
We integrate the inferred user needs (a.k.a. concepts) for each online user to our item recommender system, 
making recommendations of concept cards (one concept plus one representative item).
Two online metrics are used to measure the performance: 
click-through-rate (\textbf{CTR}) and category-discovery (\textbf{Discovery}). Detailed definitions are as follows:
\begin{equation}
\textbf{CTR} = \frac{\# \text{ concept card clicks}}{\# \text{ concept card exposes}},
\end{equation}
\begin{equation}
\textbf{Discovery} = \text{Avg}_u(\frac{\# \text{ new clk-cates in 15d}} {\# \text{ clk-cates}})
\end{equation}
where Discovery is a measurement of how many distinct categories of representative items in concept cards a user clicked today are newly discovered (not clicked in the past $15$ days in Taobao platform).
It is a temporary\footnote{Designing a proper metric to evaluate novelty in industrial recommendation is a hard and unsolved problem.} metric used in Taobao to evaluate the novelty of recommendation results.

We deploy the user needs inference module online 
and daily update our model.
When recommending a concept card, online recommender system first output a list of items as usual, 
then we pair the items in the list with inferred top concepts, and filter out those items which are not related to any concepts. In the meantime, items within top concepts will complement the list.
Followed by another ranking module, concept cards with highest scores will then be displayed to users.
\subsection{Results}

To answer the first question, we compare our model to the former strategy based on rules\footnote{Concepts are ranked by the counting number of their related items which are behaved by the user.} and the strongest baseline MCRec+ in offline setting.
Online results of A/B testing show that our model achieves highest CTR, 
which demonstrates that it can infer user needs more accurately. 
On the other hand, largest improvement on Discovery
shows our model is able to bring more novelty.

\begin{table}[th]
	\centering
	\begin{tabular}{l|c|c}
		\hline
		Strategy  & CTR   & Discovery\\
		\hline
		Rule-based  & - & - \\ 
		MCRec+  & +5.1\%   & +3.4\%\\
		Ours  & \textbf{+6.0\%} & \textbf{+5.6\%} \\

		\hline
	\end{tabular}
	\caption{Improvements on CTR and Discovery. }
	\label{tab:online}
\end{table}


To answer the second question,
we conduct a real in-app user survey on Taobao since standard metrics like CTR and Discovery may not directly represent user satisfaction.  
Due to limited resource and time, we can only finish three rounds of survey.
In each round, we randomly select 50,000 users and send them top 3 concepts inferred by model or by rule-based strategy. 
Each selected user is asked to answer a simple question: ``Are you satisfied with [X] as a recommended shopping need for you?'', where [X] is replaced by one inferred concept and the answer is YES or NO.
Around 9k users out of 50k actually answered at least one question in each round of survey.
The satisfaction rate is then calculated as the percentage of questions whose answer is YES in all answered questions, 
which is shown in \tabref{tab:survey}.
User satisfaction rate is improved by \textbf{20.6\%} if we use the proposed inference model (14.7\% for MCRec+), which demonstrates such user needs inference actually make users more satisfied.
As we can notice, the absolute number of satisfaction rate is only 41\%,
which is clearly not a large number. 
In fact, it is hard to know the true upper bound of user satisfaction rate, meaning there is ample room for us to continuously explore user needs understanding. 

\begin{table}[th]
	\centering
	\begin{tabular}{l|c|c}
		\hline
		Strategy  &  Satisfaction Rate  & Improvement\\ 
		\hline
		Rule-based  &  34\% & -\\ 
		MCRec+  &  39\% & +14.7\% \\ 
		Ours  &  \textbf{41\%} & \textbf{+20.6\%} \\
		\hline
	\end{tabular}
	\caption{Satisfaction rate from real user survey.}
	\label{tab:survey}
\end{table}

To answer the last question, 
we compare recommending a concept card with 
recommending an item (traditional item recommendation)
at the same position on ``Guess What You Like''.
Online evaluation shows significant improvements \footnote{This comparison is not entirely fair due to the different display form between a concept card and an item. 
	But the large improvements still indicate the value of user-needs driven recommendation. Integrating user needs understanding to general item recommendation is included in our future work.} of recommending concept cards:
\textbf{5.3\%} in CTR and \textbf{9.6\%} in Discovery.
If we further consider the purchases of related items in the page guided by concept cards, total sales volume (GMV) is improved by \textbf{84.0\%}, which demonstrates the great value and potential of such user-needs driven recommendation.

\subsection{Case Study}
\label{sec:case}

A major contribution of our model is that we propose a attention cube to model three-way interactions simultaneously, aiming to distinguish different importances of different factors in an e-commerce interaction, 
which may inspire us to better understand user needs.
Therefore, we analyze the attention values from several perspectives during online inference. 
All the following analysis is based on one day's user log.



During inference, if the gender of a user matches to the gender constraint of a concept, the attention weights of ``gender'' in both user profile and concept schema become nearly twice larger than not matching.
This indicates our model can explicitly learn rules such as a young female user is more likely to need a concept ``Party for girls'' rather than ``Party for boys''.

\begin{figure}[th]
	\centering
	\epsfig{file=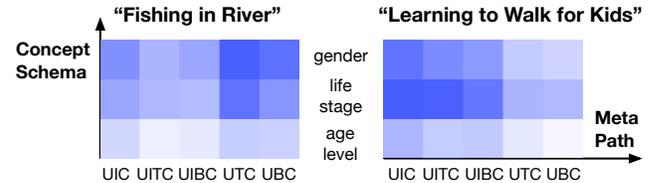, width=\columnwidth}
	\caption{Visualization of attention weights for an anonymous user. Darker colors indicate higher weights.}
	\label{fig:att2}
\end{figure}

To see if the same user has different preferences on meta-paths regarding different concepts,
 we randomly pick a user as illustrative example shown in \figref{fig:att2}.
The anonymous user has two positive interactions of concept cards: ``Learning to Walk for Kids'' and ``Fishing in River''.
After digging into transaction data,
we find that this user recently clicks a lot of kids related items, resulting high importance of behavior paths shown in his attention distribution when facing concept ``Learning to Walk for Kids''.
On the contrary, he has few behaviors related to fishing. 
Accordingly, the attention weights of preference paths are much higher than average when facing concept ``Fishing in River'' since his long-term category preference is ``fishing equipments''.

\section{Conclusion}
\label{sec:conclusion}

In this paper,
we point out that one of the biggest challenges in current e-commerce solutions
is that they are not directly driven by user needs, 
which, however, are precisely the ultimate goal of e-commerce platform try to satisfy.
To tackle it, we introduce a specially designed e-commerce knowledge graph practiced in Taobao, trying to conceptualize user needs as various shopping scenarios, also known as e-commerce concepts. 
We further proposed a deep attentive inference model to intuitively infer those concepts accurately.
On our real-world e-commerce dataset, the proposed model achieved state-of-the-art performance against several strong baselines.
After applying to online recommender system, great gain regarding both accuracy and novelty are achieved. A real user survey is conducted to demonstrate such user needs inference actually improves user satisfaction.
More importantly, we believe that the idea of conceptualizing and inferring user needs can be applied to more e-commerce applications. In the future, we will continuously explore various possibilities of ``user-needs driven'' e-commerce.

\section{Acknowledgement}
\label{sec:ack}
We deeply thank Peng Wang, Peng Yu and Guli Lin for supporting the online experiments in this paper.

\bibliographystyle{ACM-Reference-Format}
\bibliography{ref}

\end{document}